\documentclass[aps,prb,showpacs,twocolumn,groupedaddress]{revtex4}

\usepackage{amsfonts}
\usepackage{amsmath,amssymb}
\usepackage{amsthm}
\usepackage{graphicx}
\usepackage[dvips]{color}

\renewcommand{\d}{{\rm d}}

\newcommand{\imai}{{\rm i}}

\DeclareMathOperator{\Tr}{Tr}

\begin{document}

\title{Electron Transport through Nanosystems Driven by Coulomb Scattering}


\author{Benny Lassen}\altaffiliation[Now at ]{Mads Clausen Institute,
University of Southern Denmark, Grundtvigs All{\'e} 150, 
6400 S{\"o}nderborg, Denmark}\email{benny@mci.sdu.dk}
\author{Andreas Wacker}\email{Andreas.Wacker@fysik.lu.se}
\affiliation{Mathematical Physics, University of Lund, Box 118, 22100 Lund, Sweden}



\date{30. July 2007, to appear in Physical Review B}

\begin{abstract}
  Electron transmission through nanosystems is blocked if
  there are no states connecting the left and the right reservoir. 
  Electron-electron scattering can lift this blockade and we show that
  this feature can be conveniently implemented by considering a
  transport model based on many-particle states. We discuss typical
  signatures of this phenomena, such as the presence of a current
  signal for a finite bias window.
\end{abstract}


\pacs{73.23.-b,73.50.Bk,73.63.Kv}

\maketitle

\section{Introduction}
Quantum dots constitute an excellent testbed for transport through
general nanosystems, where the local density of states is dominated
by discrete localized levels. The key points are conduction
quantization \cite{WeesPRL1988} due to the discreteness
of levels, Coulomb Blockade due to electron
repulsion,\cite{MeiravPRL1990}
and the interplay between resonant
tunneling and charging in double dot
structures.\cite{WielRMP2003}
In this work we consider a further issue, the transport by
electron-electron scattering.

Electron-electron scattering is not included in
standard transmission models,\cite{LindsayAM2007} 
where the Coulomb interaction is taken into account by a
mean-field approach frequently including exchange-correlation
interactions as well. Within such models electron transport 
strongly depends on the presence of states in the system connecting both
leads.\cite{HeurichPRL2002} Here we
show that electron-electron scattering allows for
additional transport channels and that it can be 
consistently implemented using a many-particle basis following 
the concepts developed in 
Refs.~\onlinecite{KinaretPRB1992,PfannkuchePRL1995,TanakaPRB1996,HettlerPRL2003,PedersenPRB2005a,MuralidharanPRB2006}.

\begin{figure}[b]
 \centering \includegraphics[width=0.9\columnwidth]{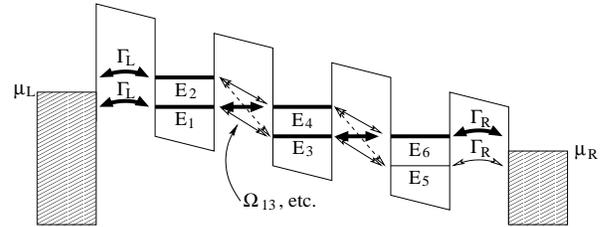}
  \caption{The triple-dot system considered. The bold lines
  represent the simplified system.}
  \label{fig:System}
\end{figure}

While in double-dot structures, each dot has direct access to a
reservoir with a continuous level density, the situation is essentially
different in triple-dot structures,\cite{GaudreauPRL2006} 
where the states in the 
central dot only couple to discrete states in the neighboring dots.
Thus the properties of these states are far more sensitive to
scattering events, which 
may essentially determine the
transport through the structure. This is precisely the situation
depicted in Fig.~\ref{fig:System}: Here the upper level 4 of
the middle dot can be filled from the left lead by resonant tunneling 
via level 1, while its lower level can be emptied  into the right lead 
by resonant tunneling via level 6. Thus the current is very sensitive
to scattering between level 4 and level 3. 
In this work we
restrict to electron-electron scattering, which is appropriate if the
phonon energies do not match the transition energy.

\section{The Model}

\subsection{The System}

The system depicted in Fig.~\ref{fig:System} is
described by the Hamiltonian
$\hat{H}=\hat{H}_\mathrm{dots}+\hat{H}_\mathrm{leads}$:
\begin{equation}
  \hat{H}_\mathrm{dots}=\sum_{i=1}^6E_i a_i^\dagger a_i
-\sum_{ij}\Omega_{ij}a_i^\dagger a_j+\hat{H}_\mathrm{ee}
\end{equation}
refers to the dot region
where $a_i^\dagger$ ($a_i$) is the creation (annihilation) operator
for the $i$'th state. Assuming that 
the states in the individual
dots are strongly localized, only states in dots next to each
other couple and we restrict to those couplings $\Omega_{ij}$ depicted in
Fig.~\ref{fig:System}.
For the Coulomb part $\hat{H}_\mathrm{ee}$ we neglect 
interactions between the leads and the dots as well as interactions 
between next-nearest neighboring dots. Then we obtain
\begin{multline}
  \hat{H}_\mathrm{ee}= U
(a_1^\dagger a_2^\dagger a_2 a_1+a_3^\dagger a_4^\dagger a_4a_3
   +a_5^\dagger a_6^\dagger a_6 a_5)\\
  +U_\mathrm{n}(a_1^\dagger a_1+a_2^\dagger a_2+a_5^\dagger a_5+a_6^\dagger
a_6)
(a_3^\dagger a_3+a_4^\dagger a_4) \\
+\{U_\mathrm{sc}a_3^\dagger a_2^\dagger a_1 a_4
    +U_\mathrm{sc}a_5^\dagger a_4^\dagger a_3 a_6+h.c.\}   \, .
\end{multline}
Here $U$ and $U_\mathrm{n}$ are the 
matrix elements of the standard Coulomb repulsion 
between states located in the same and neighboring dots, respectively.
$U_\mathrm{sc}$ describes Coulomb scattering
between different states, which is the central issue of this 
work.\cite{footnote1}

Finally, the Hamiltonian of the leads and their 
coupling to the dots reads:
\begin{equation}
\hat{H}_\mathrm{leads}=\sum_{k\ell}E_{k\ell}c_{k\ell}^\dagger
  c_{k\ell}-\sum_{ik\ell}\left( t_{ik\ell}a_i^\dagger c_{k\ell}
+t^*_{ik\ell}c^\dagger_{k\ell}a_i\right)\label{EqLeads}
\end{equation}
Here $\ell=L,R$ denotes the left and right lead, respectively.
The energies $E_{k\ell}$ in lead $\ell$
provide a continuum of states (labeled by $k$). We assume that
the corresponding density of states has the constant value $g_\ell$ in
the energy range $-W<E<W$ and is zero otherwise. Disregarding
the $k$-dependence of the tunneling matrix elements $t_{ik\ell}$ we set
$t_{1kL}=t_{2kL}=\sqrt{\Gamma_L/(2\pi g_L)}$ and
$t_{5kR}=t_{6kR}=\sqrt{\Gamma_R/(2\pi g_R)}$. These are the
transitions sketched in Fig.~\ref{fig:System}.
All other tunneling matrix elements are neglected in Eq.~(\ref{EqLeads}).
Throughout this work we restrict 
to a single spin direction for simplicity.

\subsection{Parameters}\label{SectParameters}
For specific calculations we use the parameters of 
Table~\ref{tab:Parameters} unless stated otherwise. 
\begin{table}
  \centering
  \begin{tabular}{|c|c|c|c|c|c|}
\hline
    $E_1=40$ & $E_2=60$ & $E_3=20$ & $E_4=40$ & $E_5=0$ & $E_6=20$\\ \hline
 \multicolumn{4}{|c|}{
 $\Omega_{14}=\Omega_{36}=\Omega_{23}=\Omega_{45}=0.1$} &
 \multicolumn{2}{|c|}{$\Omega_{13}=\Omega_{35}=0.05$} \\ \hline
 \multicolumn{2}{|c|}{$\Gamma_L=\Gamma_R=0.1$} & $\mu_L=50$ &
 $\mu_R=10$ & \multicolumn{2}{|c|}{$\Omega_{24}=\Omega_{46}=-0.2$}
\\ \hline
$U_\mathrm{sc}=-0.2$ & $U=10$ & $U_\mathrm{n}=3$ & $k_BT=2$ &
  \multicolumn{2}{|c|}{$W=400$} \\ \hline
\end{tabular}
\caption{Parameters used if not stated otherwise. They refer to the
  modulated nanowire discussed in 
  Sec.~\ref{SectParameters}, but have been rounded off for an easier
  recognition of scales in the plots. All energies are
  in meV.
  In the simplified system we neglect level 5 and set
  $\Omega_{13}=\Omega_{24}=\Omega_{23}=\Omega_{46}=U=U_\mathrm{n}=0$.}
\label{tab:Parameters}
\end{table}
They relate to an InAs/InP modulated nanowire structure similar 
to the structures of Refs.~\onlinecite{BjorkAPL2002a,FuhrerNL2007}.
We assume three InAs wells with a thickness of $40$~nm, which are
separated by $3$~nm thick InP barriers. The outer barriers are assumed
to be $1.5$~nm thick. 
As we are only interested in order of magnitude  estimates 
we choose the simple one-band envelope function model, with
Dirichlet boundary condition on the outside of the wire.
In addition, we assume that 
the wire is cylindrical with radius $R=20$ nm 
which enables us to reduce the problem to a 
one dimensional problem by using cylindrical coordinates, i.e.,
the single particle Hamiltonian is given by
\begin{equation}
  H=[-\frac{\partial}{\partial z}\frac{\hbar^2}{2m_{\rm
  eff}}\frac{\partial}{\partial z}+\frac{\hbar^2j_{ln}^2}{2m_{\rm
  eff}R^2}+V_c(z)]
\end{equation}
where $j_{ln}$ is the $n$'th zero of the Bessel 
function $J_l$.  $m_{\rm eff}$ is the effective mass function, 
$V_{\rm  eff}$ is the 
conduction band edge function (they are stepwise constant).
In the previous 
section we assumed that states in individual dots are strongly 
localized. One way of achieving that is to use Wannier states 
$\Psi_i(z)$ for 
individual dots, assuming a periodic repetition of the structure. 
Using the masses $m_{\rm InAs}=0.026m_e$ and $m_{\rm InP}=0.08m_e$, 
where $m_e$ is the free electron mass, and a conduction band offset of 
$0.6 eV$ \cite{BjorkAPL2002a} we get an energy difference between the 
ground state and the first excited state of $E_2-E_1=21$ meV for a given $l$ 
and $n$. The excitation energy for the radial modes
$\tfrac{\hbar^2}{2m_{\rm eff}R^2}(j_{11}^2-j_{01}^2)=29$ meV is larger
and thus these radial modes can be neglected. (In addition, the 
coupling between states of different radial symmetry should be small.)
The couplings $\Omega_{ij}$ are evaluated following Sec.~2.3 of
Ref.~\onlinecite{WackerPR2002} for a bias drop of 20 meV per period.

For the coupling to the leads we use the estimate\cite{WackerPRB1999}
\begin{equation}
  \Gamma_{iL/R}\approx \frac{2d T_i^2}{\hbar\sqrt{2(E_i-E_r)/m_{\rm InAs}}} ,
\end{equation}
where $T_i$ is the coupling element 
between Wannier state $i$ in neighboring dots for a barrier 
width of $1.5$ nm (the outer barrier) and $E_r$ is the sum of 
the conduction band edge of InAs and the radial confinement 
energy.

In general the Coulomb interaction is described by
\begin{equation}
\hat{H}_{ee}=\frac{1}{2}\sum_{ijkl} U_{ijkl}a_i^\dagger a_j^\dagger a_k a_l
\end{equation}
with 
\begin{equation}
U_{ijkl}=\int\d^3r\int\d^3r' 
\frac{e^2\quad
\varphi^*_i({\bf r})\varphi_l({\bf r})\varphi^*_j({\bf r}')\varphi_k({\bf r}')}
{4\pi\epsilon_r\epsilon_0|{\bf r}-{\bf r}'|} 
\end{equation}
Commonly, one focuses on the direct interaction of two states, where
$i=l$ and $j=k$. Taking into account the normalization of the
wave functions, we can estimate
\begin{equation}
U_{ijji}\approx \frac{e^2}{4\pi\epsilon_r\epsilon_0 d}
\end{equation}
where $d$ is the average distance between the particle densities.
Using $d\approx 10$ nm, if the states $i$ and $j$ are within the
same dot, and  $d\approx 40$ nm, if the states $i$ and $j$ are within 
adjacent dots, we obtain the values for $U$ and $U_n$ given in  
Table~\ref{tab:Parameters}, respectively, for $\epsilon_r\approx 13$.

The key scattering element $U_{\rm sc}$ corresponds to $U_{3214}$. As 
the states $\varphi^*_3({\bf r})$ and $\varphi_4({\bf r})$ [as well as
$\varphi^*_2({\bf r})$ and $\varphi_1({\bf r})$] are
orthogonal, one cannot approximate the  $1/|{\bf r}-{\bf r}'|$
potential by a constant value 
as in the case of the direct interaction discussed
above. Instead a dipole expansion is possible providing
\begin{equation}
U_{3214}\approx -\frac{e^2}{4\pi\epsilon_r\epsilon_0}\frac{2 z_{21}z_{34}}
{d^3}
\end{equation}
where $d\approx 43$ nm is the distance between the centers of
neighboring quantum dots. The $z$-matrix elements 
$z_{ij}=\int\d z \varphi^*_i(z)z\varphi_j(z)$
are evaluated for the Wannier functions, providing $z_{12}=z_{34}=-8$
nm, which gives the value in Table~\ref{tab:Parameters}.
As $U_{\rm sc}\ll U,U_n$, it is usually neglected. However, here we
show that it can have an crucial impact on the transport.


\subsection{Transport approach}

For our calculations we use a basis of many-particle states
$|a\rangle,|b\rangle,\ldots$, which diagonalize the dot Hamiltonian
$H_\mathrm{dots}$ including the Coulomb interaction.
Using the approach of Ref.~\onlinecite{PedersenPRB2005a},
but only including first-order transition processes 
between the leads and the dot region, the
following rate equations (first-order von Neumann approach, see also
Ref.~\onlinecite{PedersenPRB2007})
can be derived for the reduced density matrix of the dot
$w_{bb'}=\Tr \{\langle b|\hat{\rho}|b'\rangle\}$
where the trace is taken over all lead states $\{k\ell\}$:
\begin{equation}\begin{split} \imai \hbar \frac{\d }{\d t}&w_{bb'}=
(E_b-E_{b'})w_{bb'}  \\
&+\sum_{a,k\ell}[T_{ba}(k\ell)\phi^*_{b'a}(k\ell)
-\phi_{ba}(k\ell)T^*_{b'a}(k\ell)]\\
&+\sum_{c,k\ell}[T^*_{cb}(k\ell)\phi_{cb'}(k\ell)
-\phi^*_{cb}(k\ell)T_{cb'}(k\ell)]\, ,
\end{split}\end{equation}
with $T_{ba}(k\ell)=\sum_it_{ik\ell}\langle b| a_i^\dagger|a\rangle$ and
\begin{equation}\begin{split} 
\phi_{cb}(k\ell)=&\sum_{b'}\frac{T_{cb'}(k\ell)f_{\ell}(E_k)}
{E_k-E_c+E_b+\imai 0^+}w_{b'b}\\
&-\sum_{c'}\frac{T_{c'b}(k\ell)[1-f_{\ell}(E_k)]}
{E_k-E_c+E_b+\imai 0^+}w_{cc'} 
\, .
\label{EqPhi}
\end{split}\end{equation}
Here $f_{\ell}(E)=(1+e^{(E-\mu_\ell)/k_bT})^{-1}$ 
is the Fermi distribution for the 
lead $\ell$ with electrochemical potential $\mu_\ell$.
The current from lead $\ell$ into the sample is given 
by $J_{\ell}=\sum_{cb}J_{\ell}(cb)$, where
\begin{equation}
J_{\ell}(cb)=-e\frac{2}{\hbar}\Im\left\{\sum_{k}T^*_{cb}(k\ell)
\phi_{cb}(k\ell)\right\} 
\label{EqJLcb}
\end{equation}
is the part of the  current associated with transitions between states 
$b$ and $c$ within the dot.\cite{PedersenPRB2005a,PedersenPRB2007} 
We disregard
the sign of the electron charge $e$, so that the sign of the electrical current
equals the sign of the particle current.

It should be noted that we obtain the Pauli Master 
Equation, \cite{BeenakkerPRB1991} if we neglect the off-diagonal elements of
the density matrix. However,
this approximation is only reasonable as long as the spacing between
the many-particle energies is large compared to the
contact couplings $\Gamma$.\cite{PedersenPRB2007}
This is not the case for the systems considered here,
so the full set of equations is needed.


\section{Results for the simplified system}

At first we study a simplified
system where we neglect state 5 and all interdot tunneling processes
except for $\Omega_{23}$ and $\Omega_{36}$ which are in resonance.
This corresponds to the thick lines and arrows in Fig.~\ref{fig:System}.
In order to avoid complications due to Coulomb charging, we set 
$U_\mathrm{n}=U=0$, thus focusing on the scattering 
via $U_\mathrm{sc}$.

\begin{figure}
 \includegraphics[width=0.8\columnwidth]{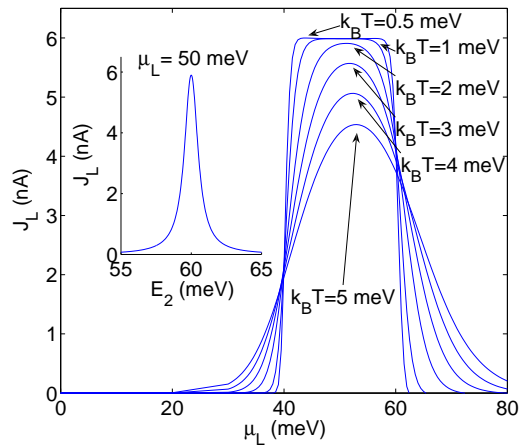}
  \caption{Currents as a function of $\mu_L$ and $E_2$ (subfigure).
  Other parameters from  Table~\ref{tab:Parameters} (simplified
  system).}
  \label{fig:IV1}
\end{figure}

In Fig.~\ref{fig:IV1} we show the current as a function of the left
Fermi level $\mu_L$.
There is no current until the left Fermi
level comes within the vicinity of the ground-state
of the first dot ($E_1=40$ meV). At this point electrons 
start to flow from the left lead into this state
and further into the excited state of the second dot. 
If both states 1 and 4 are occupied, 
the Coulomb scattering via $U_\mathrm{sc}$ is possible. 
This process transfers one electron from level 1 
to level 2 and a second electron from level 4 to level 3, 
which can subsequently reach the right lead
via level 6. Thus Coulomb scattering establishes a transport path 
through the nanosystem.
However, when the left Fermi level comes
into the vicinity of the excited state of the first dot ($E_2=60$ meV) 
electrons will start to occupy this state. This causes a 
decrease in the current (see Fig.~\ref{fig:IV1}) as Pauli blocking hinders
the scattering process addressed above.
Likewise the temperature
dependence essentially follows the  probability
\begin{equation}
F=f_L(E_1)f_L(E_4)[1-f_L(E_2)][1-f_R(E_3)]
\end{equation} 
to find states 
$1$ and $4$ occupied while states $2$ and $3$ are empty.
The relevance of level $E_2$ for the transport is further demonstrated in
the subfigure of  Fig.~\ref{fig:IV1}, showing that current only flows
through the triple-dot structure if $E_2-E_1\approx E_4-E_3$, where
the Coulomb scattering is energetically allowed.

This presence of current 
enhancement in a finite bias window $\Delta \mu_L$ matching
the energy transfer $\Delta E$ by the scattering process 
is the characteristic signal of 
electron transport by Coulomb scattering. This Pauli-blocking of the 
scattering from level 4 to level 3 by occupation of the 
further level 2 does not appear
for other inelastic scattering mechanisms such as phonon
scattering.


\section{Description by scattering}

\begin{figure}[t]
 \includegraphics[width=0.8\columnwidth]{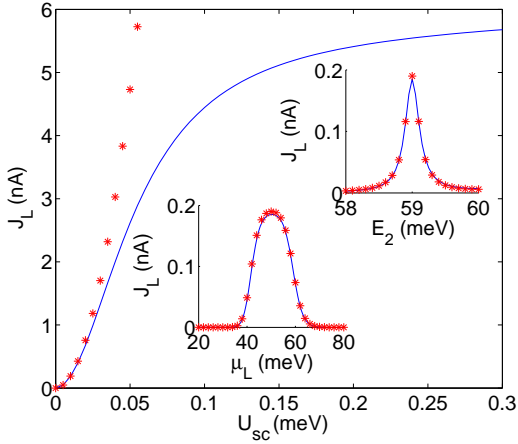}
  \caption[a]{Current as a function of $U_\mathrm{sc}$ with
  $\Omega_{14}=\Omega_{36}=1$~meV and $E_2=59$ meV. 
  The two subfigures show the current
  for $U_\mathrm{sc}=0.01$~meV as a function of $\mu_L$ and $E_2$,
  respectively. Other parameters from  
  Table~\ref{tab:Parameters} (simplified system).
  The solid lines are calculated by the
  first-order von Neumann approach
and the dots depict $J_L=eR_\mathrm{tr}$, 
where $R_\mathrm{tr}$ is the transition 
rate, Eq.~(\ref{eq:FGR}), for the electron-electron scattering process.}
  \label{fig:IU1}
\end{figure}

The description based on scattering given above
becomes quantitative if the Coulomb scattering is the limiting
process for transport through the device, i.e., if $U_\mathrm{sc}$ is
significantly smaller than $\Omega_{14}$ and $\Omega_{36}$.
For this reason, we have performed calculations for the 
increased values $\Omega_{14}=\Omega_{36}=1$~meV, see Fig.~\ref{fig:IU1}. 
The strong coupling
between the states $3$ and $6$ yields a bonding and an anti-bonding
state, with energies $20\mp 1$ meV. Therefore the resonance condition
for Coulomb scattering is now satisfied at 
$E_2=59$ meV and  $E_2=61$ meV (not shown) as displayed 
in the right subfigure of Fig.~\ref{fig:IU1}. 

Fermi's golden rule provides us with the transition rate by
Coulomb scattering into the anti-bonding state between 3 and 6:
\begin{equation}\label{eq:FGR}
\begin{split}
R_\mathrm{tr}=&\frac{U_\mathrm{sc}^2}{2\hbar}\frac{\Gamma_\mathrm{eff}}
{(E_2+E_3+\Omega_{36}-E_1-E_4)^2+\Gamma_\mathrm{eff}^2/4}F
\end{split}
\end{equation}
Here we have replaced the energy-conserving $\delta$-function by
a Lorentzian, representing life-time broadening due to
the coupling to leads. $\Gamma_\mathrm{eff}=2\Gamma_L+\Gamma_R/2$ 
is the sum of broadenings for the individual states: 
$\Gamma_L$ for the levels 1 and 2, and
$\Gamma_R/2$ for the anti-bonding combination of 3 and 6.
Fig.~\ref{fig:IU1} shows that Fermi's golden rule provides a full
quantitative description for small $U_\mathrm{sc}$.
However, for larger values of $U_\mathrm{sc}$, this simple
reasoning, based on single-particle states, fails. 
In particular, the width of the current peak becomes much
broader than the simple life-time broadening $\Gamma_\mathrm{eff}$
(see subfigure of  Fig.~\ref{fig:IV1}), which makes it easier to
observe the effect in a real system with imprecise control over the level 
energies.


\section{Description by many-particle states}

\begin{figure}
 \includegraphics[width=0.95\columnwidth]{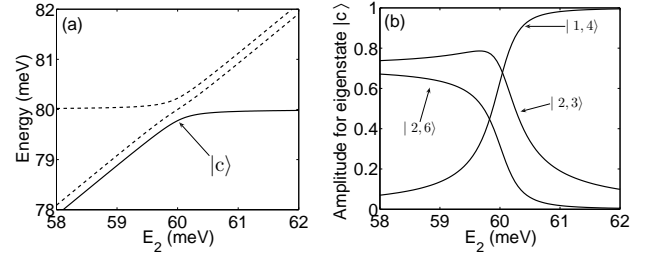}
  \caption[a]{a) Energies of selected two-particle states as a function of 
  $E_2$. b) Coefficients of state $|c\rangle$. Other parameters from Table
  \ref{tab:Parameters} (simplified system).}
  \label{fig:Disp}
\end{figure}

Now we want to sketch, how this scattering-induced transport emerges
within a basis of many-particle states, which takes into account 
the entire Coulomb interaction.
For the parameters of Table \ref{tab:Parameters} the 
anti-symmetrized  two-particle 
product states $|1,4\rangle$, $|2,3\rangle$, and  $|2,6\rangle$
all have the same sum of single-particle energies $E_i+E_j=80$~meV.
They couple to each other due to the matrix elements $U_\mathrm{sc}$ and 
$\Omega_{36}$, resulting in the three many-particle states depicted in
Fig.~\ref{fig:Disp}. For $E_2\approx 60$~meV, the three states are
highly entangled and we focus in the following on one of these
entangled states, denoted by $|c\rangle$. This state contributes to a circle
of transitions between different many-particle states depicted in 
Fig.~\ref{fig:TransportProcess}:
\begin{figure}
  \includegraphics[width=0.9\columnwidth]{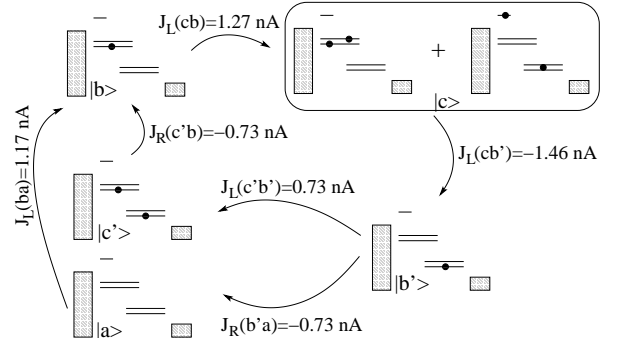}
  \caption{Diagrammatic representation of a circle of transport
  processes involving the many-particle states  $|b\rangle$, 
  $|c\rangle$, $|b'\rangle$, $|a\rangle$, and $|c\rangle$.
  The two sets of double lines are the bonding and
  anti-bonding combinations of the states $1$, $4$ and $3$, $6$,
  respectively.}
  \label{fig:TransportProcess}
\end{figure}
The state $|c\rangle$ can be reached by tunneling of an electron from
the left lead into the state $|b\rangle$ (process from upper left to
upper right). Here $|b\rangle$ is the binding one-particle state
combining levels 1 and 4.  By removing an electron towards the left
lead (at a higher energy than before),  the state $|c\rangle$ decays
to the one-particle state  $|b'\rangle$, the  binding state combining
levels 3 and 6. Then the original state $|b\rangle$ is restored by one
electron  tunneling from the state $|b'\rangle$ to the right lead and
one electron tunneling into the state $|b\rangle$ from the left lead,
which can happen in two different sequential orders.  The key issue
for the existence of this circle  is the presence of the entangled
state $|c\rangle$, which enables the transition between $|b\rangle$ and
$|b'\rangle$ via two single-electron tunneling processes. Therefore
the current drops, if the product states  $|1,4\rangle$ and
$(|2,3\rangle+|2,6\rangle)/\sqrt{2}$ are detuned by varying $E_2$.

The currents $J_{L/R}(cb)$ in Fig.~\ref{fig:TransportProcess} denote 
the contribution of the transitions $b \leftrightarrow c$ 
between the corresponding many-particle states to the current from the
left/right lead into the system, respectively, as given in 
Eq.~(\ref{EqJLcb}). The magnitude of
these currents corresponds to the transition rate between the  states.
Fig.~\ref{fig:TransportProcess} shows that the ingoing and outgoing
rates partially balance for all states depicted. Nevertheless, there are plenty
of further transitions, which make the full picture far more involved.
In total this circle provides 
$J_L=1.71$ nA and $J_R=-1.46$ nA, which constitutes only a part of the total
current $J_L=-J_R=5.9$ nA. The remaining part is
carried by similar circles involving the other many-particle states as
well as more complicated transitions which cannot be separated into
circles that easily.

Finally, note that the electrons enter the structure with 
$E\approx E_1$ from the left contact and leave the structure with 
$E\approx E_5$ to the right contact as well as with $E\approx E_2$ 
to the left contact. Thus there is no single transmission channel at 
a given energy as typical for the frequently used transmission models.

\section{Results for the full system}

\begin{figure}
\includegraphics[angle=0, width=0.8\columnwidth]{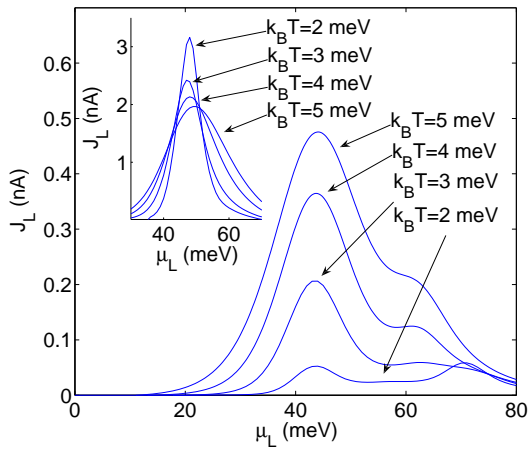}
  \caption{Current as a function of $\mu_L$ for different temperatures. 
Other parameters from  Table~\ref{tab:Parameters}.
In the subfigure the single-particle energies
  $E_1=43$~meV,
  $E_2=63$~meV,
  $E_3=20$~meV, $E_4=40$~meV, $E_5=-4$~meV, and $E_6=16$~meV are
used.}
  \label{fig:IVTFull}
\end{figure}

In Fig.~\ref{fig:IVTFull}
we present results for the full system, i.e., using all parameters given in 
Table~\ref{tab:Parameters}.
Like in Fig.~\ref{fig:IV1} we observe a current signal 
in a finite region of $\mu_L$, which is a key feature of the current 
induced by electron-electron scattering (the current is below 
0.006 nA if $U_\mathrm{sc}=0$ is used).
However, contrary to the simplified system 
in Fig.~\ref{fig:IV1}, the peak current
increases with temperature for the full system and is much 
weaker. This is due to the presence of an electron 
in state $5$ which breaks the alignment between the
levels 3 and 6 by Coulomb repulsion. With increasing temperature,
the probability for state 5 to be empty increases and so does the current. 
In the subfigure of Fig.~\ref{fig:IVTFull} we show 
results for a case where the single-particle energies have been 
modified to compensate for  charging effects, which provides results
similar to Fig.~\ref{fig:IV1}. In both cases we observe
enhanced current in (multiple) finite bias windows $\Delta \mu_L$ matching
the energy transfer $\Delta E$, which are however smeared out by
temperature.
This shows that the essential features
of transport by electron-electron scattering are robust
with respect to other electron-electron interaction mechanisms.


\section{Conclusion}

We have shown that Coulomb scattering provides a current channel
for transport through a triple-dot system. The mechanism
holds for general nanosystems exhibiting two pairs of states with
a similar level spacing $\Delta E$. An example is the 
conduction through a macro molecule, where an appropriate chemical 
group appears twice. A typical signature is a current
signal for a finite bias window, matching the energy transfer $\Delta E$.
If the Coulomb scattering is the slowest transfer process involved, 
a simple description based on Fermi's golden rule is valid. 
Otherwise a systematic implementation is
possible within a basis of many-particle states, 
which reflects the total Coulomb interaction for the
nanosystem.

\acknowledgments
We thank J.~N.~Pedersen for helpful discussion.
This work was supported by Villum Kann Rasmussen fonden and
the Swedish Research Council (VR).


\begin{thebibliography}{19}
\expandafter\ifx\csname natexlab\endcsname\relax\def\natexlab#1{#1}\fi
\expandafter\ifx\csname bibnamefont\endcsname\relax
  \def\bibnamefont#1{#1}\fi
\expandafter\ifx\csname bibfnamefont\endcsname\relax
  \def\bibfnamefont#1{#1}\fi
\expandafter\ifx\csname citenamefont\endcsname\relax
  \def\citenamefont#1{#1}\fi
\expandafter\ifx\csname url\endcsname\relax
  \def\url#1{\texttt{#1}}\fi
\expandafter\ifx\csname urlprefix\endcsname\relax\def\urlprefix{URL }\fi
\providecommand{\bibinfo}[2]{#2}
\providecommand{\eprint}[2][]{\url{#2}}

\bibitem[{\citenamefont{van Wees et~al.}(1988)\citenamefont{van Wees, van
  Houten, Beenakker, Williamson, Kouwenhoven, van~der Marel, and
  Foxon}}]{WeesPRL1988}
\bibinfo{author}{\bibfnamefont{B.~J.} \bibnamefont{van Wees}},
  \bibinfo{author}{\bibfnamefont{H.}~\bibnamefont{van Houten}},
  \bibinfo{author}{\bibfnamefont{C.~W.~J.} \bibnamefont{Beenakker}},
  \bibinfo{author}{\bibfnamefont{J.~G.} \bibnamefont{Williamson}},
  \bibinfo{author}{\bibfnamefont{L.~P.} \bibnamefont{Kouwenhoven}},
  \bibinfo{author}{\bibfnamefont{D.}~\bibnamefont{van~der Marel}},
  \bibnamefont{and} \bibinfo{author}{\bibfnamefont{C.~T.} \bibnamefont{Foxon}},
  \bibinfo{journal}{Phys. Rev. Lett.} \textbf{\bibinfo{volume}{60}},
  \bibinfo{pages}{848} (\bibinfo{year}{1988}).

\bibitem[{\citenamefont{Meirav et~al.}(1990)\citenamefont{Meirav, Kastner, and
  Wind}}]{MeiravPRL1990}
\bibinfo{author}{\bibfnamefont{U.}~\bibnamefont{Meirav}},
  \bibinfo{author}{\bibfnamefont{M.~A.} \bibnamefont{Kastner}},
  \bibnamefont{and} \bibinfo{author}{\bibfnamefont{S.~J.} \bibnamefont{Wind}},
  \bibinfo{journal}{Phys.~Rev.~Lett.} \textbf{\bibinfo{volume}{65}},
  \bibinfo{pages}{771} (\bibinfo{year}{1990}).

\bibitem[{\citenamefont{van~der Wiel et~al.}(2003)\citenamefont{van~der Wiel,
  De~Franceschi, Elzerman, Fujisawa, Tarucha, and Kouwenhoven}}]{WielRMP2003}
\bibinfo{author}{\bibfnamefont{W.~G.} \bibnamefont{van~der Wiel}},
  \bibinfo{author}{\bibfnamefont{S.}~\bibnamefont{De~Franceschi}},
  \bibinfo{author}{\bibfnamefont{J.~M.} \bibnamefont{Elzerman}},
  \bibinfo{author}{\bibfnamefont{T.}~\bibnamefont{Fujisawa}},
  \bibinfo{author}{\bibfnamefont{S.}~\bibnamefont{Tarucha}}, \bibnamefont{and}
  \bibinfo{author}{\bibfnamefont{L.~P.} \bibnamefont{Kouwenhoven}},
  \bibinfo{journal}{Rev. Mod. Phys.} \textbf{\bibinfo{volume}{75}},
  \bibinfo{pages}{1} (\bibinfo{year}{2003}).

\bibitem[{\citenamefont{Lindsay and Ratner}(2007)}]{LindsayAM2007}
\bibinfo{author}{\bibfnamefont{S.~M.} \bibnamefont{Lindsay}} \bibnamefont{and}
  \bibinfo{author}{\bibfnamefont{M.~A.} \bibnamefont{Ratner}},
  \bibinfo{journal}{Advanced Materials} \textbf{\bibinfo{volume}{19}},
  \bibinfo{pages}{23} (\bibinfo{year}{2007}), \bibinfo{note}{and references
  cited therein}.

\bibitem[{\citenamefont{Heurich et~al.}(2002)\citenamefont{Heurich, Cuevas,
  Wenzel, and Sch{\"o}n}}]{HeurichPRL2002}
\bibinfo{author}{\bibfnamefont{J.}~\bibnamefont{Heurich}},
  \bibinfo{author}{\bibfnamefont{J.~C.} \bibnamefont{Cuevas}},
  \bibinfo{author}{\bibfnamefont{W.}~\bibnamefont{Wenzel}}, \bibnamefont{and}
  \bibinfo{author}{\bibfnamefont{G.}~\bibnamefont{Sch{\"o}n}},
  \bibinfo{journal}{Phys.~Rev.~Lett.} \textbf{\bibinfo{volume}{88}},
  \bibinfo{pages}{256803} (\bibinfo{year}{2002}).

\bibitem[{\citenamefont{Kinaret et~al.}(1992)\citenamefont{Kinaret, Meir,
  Wingreen, Lee, and Wen}}]{KinaretPRB1992}
\bibinfo{author}{\bibfnamefont{J.~M.} \bibnamefont{Kinaret}},
  \bibinfo{author}{\bibfnamefont{Y.}~\bibnamefont{Meir}},
  \bibinfo{author}{\bibfnamefont{N.~S.} \bibnamefont{Wingreen}},
  \bibinfo{author}{\bibfnamefont{P.~A.} \bibnamefont{Lee}}, \bibnamefont{and}
  \bibinfo{author}{\bibfnamefont{X.-G.} \bibnamefont{Wen}},
  \bibinfo{journal}{Phys.~Rev.~B} \textbf{\bibinfo{volume}{46}},
  \bibinfo{pages}{4681} (\bibinfo{year}{1992}).

\bibitem[{\citenamefont{Pfannkuche and Ulloa}(1995)}]{PfannkuchePRL1995}
\bibinfo{author}{\bibfnamefont{D.}~\bibnamefont{Pfannkuche}} \bibnamefont{and}
  \bibinfo{author}{\bibfnamefont{S.~E.} \bibnamefont{Ulloa}},
  \bibinfo{journal}{Phys.~Rev.~Lett.} \textbf{\bibinfo{volume}{74}},
  \bibinfo{pages}{1194} (\bibinfo{year}{1995}).

\bibitem[{\citenamefont{Tanaka and Akera}(1996)}]{TanakaPRB1996}
\bibinfo{author}{\bibfnamefont{Y.}~\bibnamefont{Tanaka}} \bibnamefont{and}
  \bibinfo{author}{\bibfnamefont{H.}~\bibnamefont{Akera}},
  \bibinfo{journal}{Phys.~Rev.~B} \textbf{\bibinfo{volume}{53}},
  \bibinfo{pages}{3901} (\bibinfo{year}{1996}).

\bibitem[{\citenamefont{Hettler et~al.}(2003)\citenamefont{Hettler, Wenzel,
  Wegewijs, and Schoeller}}]{HettlerPRL2003}
\bibinfo{author}{\bibfnamefont{M.~H.} \bibnamefont{Hettler}},
  \bibinfo{author}{\bibfnamefont{W.}~\bibnamefont{Wenzel}},
  \bibinfo{author}{\bibfnamefont{M.~R.} \bibnamefont{Wegewijs}},
  \bibnamefont{and}
  \bibinfo{author}{\bibfnamefont{H.}~\bibnamefont{Schoeller}},
  \bibinfo{journal}{Phys.~Rev.~Lett.} \textbf{\bibinfo{volume}{90}},
  \bibinfo{pages}{076805} (\bibinfo{year}{2003}).

\bibitem[{\citenamefont{Pedersen and Wacker}(2005)}]{PedersenPRB2005a}
\bibinfo{author}{\bibfnamefont{J.~N.} \bibnamefont{Pedersen}} \bibnamefont{and}
  \bibinfo{author}{\bibfnamefont{A.}~\bibnamefont{Wacker}},
  \bibinfo{journal}{Phys.~Rev.~B} \textbf{\bibinfo{volume}{72}},
  \bibinfo{pages}{195330} (\bibinfo{year}{2005}).

\bibitem[{\citenamefont{Muralidharan et~al.}(2006)\citenamefont{Muralidharan,
  Ghosh, and Datta}}]{MuralidharanPRB2006}
\bibinfo{author}{\bibfnamefont{B.}~\bibnamefont{Muralidharan}},
  \bibinfo{author}{\bibfnamefont{A.~W.} \bibnamefont{Ghosh}}, \bibnamefont{and}
  \bibinfo{author}{\bibfnamefont{S.}~\bibnamefont{Datta}},
  \bibinfo{journal}{Phys.~Rev.~B} \textbf{\bibinfo{volume}{73}},
  \bibinfo{pages}{155410} (\bibinfo{year}{2006}).

\bibitem[{\citenamefont{Gaudreau et~al.}(2006)\citenamefont{Gaudreau,
  Studenikin, Sachrajda, Zawadzki, Kam, Lapointe, Korkusinski, and
  Hawrylak}}]{GaudreauPRL2006}
\bibinfo{author}{\bibfnamefont{L.}~\bibnamefont{Gaudreau}},
  \bibinfo{author}{\bibfnamefont{S.~A.} \bibnamefont{Studenikin}},
  \bibinfo{author}{\bibfnamefont{A.~S.} \bibnamefont{Sachrajda}},
  \bibinfo{author}{\bibfnamefont{P.}~\bibnamefont{Zawadzki}},
  \bibinfo{author}{\bibfnamefont{A.}~\bibnamefont{Kam}},
  \bibinfo{author}{\bibfnamefont{J.}~\bibnamefont{Lapointe}},
  \bibinfo{author}{\bibfnamefont{M.}~\bibnamefont{Korkusinski}},
  \bibnamefont{and} \bibinfo{author}{\bibfnamefont{P.}~\bibnamefont{Hawrylak}},
  \bibinfo{journal}{Phys.~Rev.~Lett.} \textbf{\bibinfo{volume}{97}},
  \bibinfo{pages}{036807} (\bibinfo{year}{2006}).

\bibitem[{foo()}]{footnote1}
\bibinfo{note}{Further terms like $a_3^\dagger a_1^\dagger a_1 a_4$ have been
  neglected here as they are never in resonance. They are, however, of
  relevance to fully restore the locality of scattering.}

\bibitem[{\citenamefont{Bj{\"o}rk et~al.}(2002)\citenamefont{Bj{\"o}rk,
  Ohlsson, Sass, Persson, Thelander, Magnusson, Deppert, Wallenberg, and
  Samuelson}}]{BjorkAPL2002a}
\bibinfo{author}{\bibfnamefont{M.~T.} \bibnamefont{Bj{\"o}rk}},
  \bibinfo{author}{\bibfnamefont{B.~J.} \bibnamefont{Ohlsson}},
  \bibinfo{author}{\bibfnamefont{T.}~\bibnamefont{Sass}},
  \bibinfo{author}{\bibfnamefont{A.~I.} \bibnamefont{Persson}},
  \bibinfo{author}{\bibfnamefont{C.}~\bibnamefont{Thelander}},
  \bibinfo{author}{\bibfnamefont{M.~H.} \bibnamefont{Magnusson}},
  \bibinfo{author}{\bibfnamefont{K.}~\bibnamefont{Deppert}},
  \bibinfo{author}{\bibfnamefont{L.~R.} \bibnamefont{Wallenberg}},
  \bibnamefont{and}
  \bibinfo{author}{\bibfnamefont{L.}~\bibnamefont{Samuelson}},
  \bibinfo{journal}{Appl.~Phys.~Lett.} \textbf{\bibinfo{volume}{80}},
  \bibinfo{pages}{1058} (\bibinfo{year}{2002}).

\bibitem[{\citenamefont{Fuhrer et~al.}(2007)\citenamefont{Fuhrer, Fr{\"o}berg,
  Pedersen, Larsson, Wacker, Pistol, and Samuelson}}]{FuhrerNL2007}
\bibinfo{author}{\bibfnamefont{A.}~\bibnamefont{Fuhrer}},
  \bibinfo{author}{\bibfnamefont{L.~E.} \bibnamefont{Fr{\"o}berg}},
  \bibinfo{author}{\bibfnamefont{J.~N.} \bibnamefont{Pedersen}},
  \bibinfo{author}{\bibfnamefont{M.~W.} \bibnamefont{Larsson}},
  \bibinfo{author}{\bibfnamefont{A.}~\bibnamefont{Wacker}},
  \bibinfo{author}{\bibfnamefont{M.-E.} \bibnamefont{Pistol}},
  \bibnamefont{and}
  \bibinfo{author}{\bibfnamefont{L.}~\bibnamefont{Samuelson}},
  \bibinfo{journal}{Nano Letters} \textbf{\bibinfo{volume}{7}},
  \bibinfo{pages}{243} (\bibinfo{year}{2007}).

\bibitem[{\citenamefont{Wacker}(2002)}]{WackerPR2002}
\bibinfo{author}{\bibfnamefont{A.}~\bibnamefont{Wacker}},
  \bibinfo{journal}{Phys.~Rep.} \textbf{\bibinfo{volume}{357}},
  \bibinfo{pages}{1} (\bibinfo{year}{2002}).

\bibitem[{\citenamefont{Wacker and Hu}(1999)}]{WackerPRB1999}
\bibinfo{author}{\bibfnamefont{A.}~\bibnamefont{Wacker}} \bibnamefont{and}
  \bibinfo{author}{\bibfnamefont{B.~Y.-K.} \bibnamefont{Hu}},
  \bibinfo{journal}{Phys.~Rev.~B} \textbf{\bibinfo{volume}{60}},
  \bibinfo{pages}{16039} (\bibinfo{year}{1999}).

\bibitem[{\citenamefont{Pedersen et~al.}(2007)\citenamefont{Pedersen, Lassen,
  Wacker, and Hettler}}]{PedersenPRB2007}
\bibinfo{author}{\bibfnamefont{J.~N.} \bibnamefont{Pedersen}},
  \bibinfo{author}{\bibfnamefont{B.}~\bibnamefont{Lassen}},
  \bibinfo{author}{\bibfnamefont{A.}~\bibnamefont{Wacker}}, \bibnamefont{and}
  \bibinfo{author}{\bibfnamefont{M.~H.} \bibnamefont{Hettler}},
  \bibinfo{journal}{Phys.~Rev.~B} \textbf{\bibinfo{volume}{75}},
  \bibinfo{pages}{235314} (\bibinfo{year}{2007}).

\bibitem[{\citenamefont{Beenakker}(1991)}]{BeenakkerPRB1991}
\bibinfo{author}{\bibfnamefont{C.~W.~J.} \bibnamefont{Beenakker}},
  \bibinfo{journal}{Phys.~Rev.~B} \textbf{\bibinfo{volume}{44}},
  \bibinfo{pages}{1646} (\bibinfo{year}{1991}).

\end{thebibliography}

\end{document}